\documentclass[twocolumn,aps,prl,showpacs,preprintnumbers,amsmath,amssymb,groupedaddress,superscriptaddress]{revtex4}

\usepackage{graphicx}
\usepackage{dcolumn}
\usepackage{bm}

\begin{document}
\def\e{\mathcal{E}}
\title{Optimal light storage with full pulse shape control}

\author{Irina Novikova}
\affiliation{Department of Physics, College of William \textsc{\&} Mary,
Williamsburg, Virginia 23185, USA}
\author{Nathaniel B. Phillips}
\affiliation{Department of Physics, College of William \textsc{\&} Mary,
Williamsburg, Virginia 23185, USA}
\author{Alexey V. Gorshkov}
\affiliation{Department of Physics, Harvard University, Cambridge, Massachusetts 02138, USA}

\date{\today}

\begin{abstract}

We experimentally demonstrate optimal storage and retrieval of light pulses of
arbitrary shape in atomic ensembles. 
By shaping auxiliary control pulses, we attain efficiencies approaching the fundamental limit 
and achieve precise retrieval into any predetermined temporal profile.
Our techniques, demonstrated in warm Rb vapor, are applicable to a wide range of systems and protocols. As an example,
we present their potential application to the creation of optical time-bin qubits and to controlled
partial retrieval.


\end{abstract}


%
%

\maketitle


Quantum memory for light is essential for the implementation of long-distance
quantum communication \cite{DLCZ} and of linear optical quantum computation
\cite{kok07}. Both applications put forth two important requirements for the quantum
memory: (i) the memory efficiency  is high (\emph{i.e.}, the probability of losing a photon during storage and retrieval
is low) and (ii) the retrieved photonic wavepacket has a well-controlled shape
to enable interference with other photons. In this Letter, we report on the
first experimental demonstration of this full optimal control over light
storage and retrieval: by shaping an auxiliary control field, we
store an arbitrary incoming signal pulse shape and then retrieve
it into any desired wavepacket with the
maximum efficiency possible for the given memory. 
While our results are obtained in warm Rb vapor using electromagnetically
induced transparency (EIT) \cite{lukin03rmp,eisaman05}, the presented procedure
is universal~\cite{gorshkovPRL} and applicable to a wide range of systems,
including ensembles of cold atoms \cite{choi08, kuzmich05, simon07} and
solid-state impurities  \cite{alexander06, staudt07}, as well as to other light
storage protocols (\emph{e.g.}, the far-off-resonant Raman scheme
\cite{kozhekin00}).

\begin{figure}
\includegraphics[width=1.0\columnwidth]{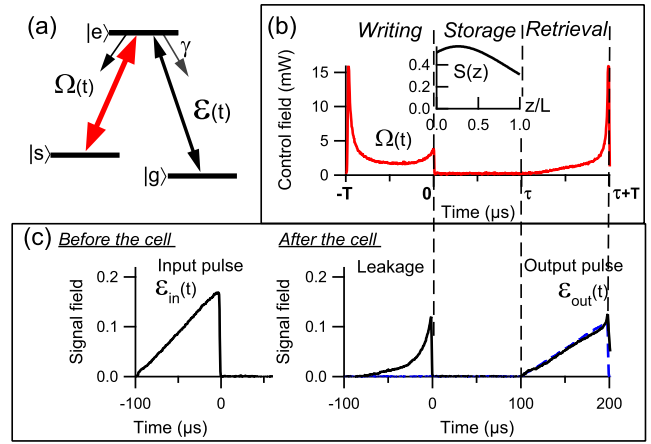}%
\caption {(Color online) (a) Schematic of the three-level $\Lambda$ interaction
scheme. Control~(b) and signal~(c) fields in pulse-shape-preserving storage of
a 
``positive-ramp" pulse using a calculated optimal control field envelope
$\Omega(t)$. During the \emph{writing} stage ($t < 0$), the input pulse
$\e_\textrm{in}(t)$ is mapped onto the optimal spin-wave $S(z)$
[inset in (b)],
while a fraction of the pulse escapes the cell (leakage). 
After a \emph{storage} time $\tau$, the spin-wave $S(z)$ is mapped into an
output signal pulse $\e_\textrm{out}(t)$ during the \emph{retrieval} stage. The
dashed blue line in (c) shows the target output pulse shape.
\label{basicstorage.fig}}
\end{figure}

We consider the propagation of a weak signal pulse in the
presence of a strong classical control field in a resonant $\Lambda$-type
ensemble under EIT conditions, as shown in Fig.~\ref{basicstorage.fig}(a). An
incoming signal pulse propagates with slow group velocity $v_g$, which is
uniform throughout the medium and is proportional to the intensity of the
control field $v_\textrm{g} \approx 2 |\Omega|^2/(\alpha \gamma)\ll
c$~\cite{fleischhauer}. Here, $\Omega$ is control Rabi frequency, $\gamma$ is
the decay rate of the optical polarization, and $\alpha$ is the
absorption coefficient, 
so that $\alpha L$ is the optical depth of the atomic medium of length $L$. For
quantum memory applications, a signal pulse can be ``stored'' (\emph{i.e.}\
reversibly mapped) onto a collective spin excitation of an ensemble (spin wave)
by reducing the control intensity to zero~\cite{fleischhauer}. In the limit of
infinitely large optical depth and negligible ground state decoherence, any
input pulse can be converted into a spin wave and back with 100\% efficiency,
satisfying requirement (i). Under the same conditions, any desired
output pulse shape can be easily obtained 
by adjusting the control field power (and hence the group velocity) as the
pulse exists the medium, in accordance with requirement (ii).
However, most current experimental realizations of
ensemble-based quantum memories 
operate at limited optical depth
$\alpha L \lesssim 10$ due to various constraints~\cite{kuzmich05,
choi08,simon07,eisaman05,alexander06, staudt07}. At finite $\alpha L$, losses limit the
maximum achievable memory efficiency 
to a value below 100\%, making efficiency optimization and output-pulse shaping important and nontrivial~\cite{gorshkovPRL}.


In this Letter, we experimentally demonstrate the capability to satisfy both
quantum memory requirements in an ensemble with a limited optical depth. 
Specifically, by adjusting the
control field envelopes for several arbitrarily selected input pulse shapes, we
demonstrate precise retrieval into any desired output pulse shape with
experimental memory efficiency very close to the fundamental
limit~\cite{gorshkovPRL, gorshkovPRA2}. This ability to achieve maximum
efficiency for any input pulse shape is crucial
when optimization with respect to the input wavepacket \cite{novikovaPRLopt} is
not applicable  (\emph{e.g.}, if the photons are generated by parametric
down-conversion \cite{neergaard07}).
At the same time, 
 control over the outgoing mode, with precision far beyond the early attempts~\cite{hau01,eisaman04,patnaik04} is essential for experiments based on
the interference of photons stored under different experimental conditions
(\emph{e.g.},\ in atomic ensembles with different optical depths), or stored a
different number of times. 
In addition, control over output pulse duration may
also allow one to reduce sensitivity to noise (\emph{e.g.}, jitter).
It is important to note that although our experiment
used weak classical pulses, the linearity of the corresponding equations of
motion \cite{gorshkovPRL} ensures direct applicability of our results to
quantum states 
confined to the mode defined by
the classical pulse.


The experimental setup is described in detail in Ref.~\cite{phillipsPRA08}. We
phase-modulated the output of an  external cavity diode laser to produce
modulation sidebands separated by the ground-state hyperfine splitting of
$^{87}$Rb ($\Delta_\mathrm{HF}=6.835$~GHz). For this experiment, we tuned the
zeroth order (control field) to the $F=2 \rightarrow F'=2$ transition of the
$^{87}$Rb $\mathrm{D}_1$ line, while the $+1$ modulation sideband played the
role of the signal field and was tuned to the $F=1 \rightarrow F'=2$
transition. The amplitudes of the control and signal fields were controlled
independently by simultaneously adjusting the phase modulation amplitude (by
changing the rf power sent to the electro-optical modulator) and the total
intensity in the laser beam (using an acousto-optical modulator). Typical peak
control field and signal field powers 
were 18~mW and
50~$\mu$W, respectively. The $-1$ modulation sideband was
 suppressed to $10\%$ of its original intensity using a temperature-tunable
 Fabri-Perot etalon.
In the experiment, we used a cylindrical Pyrex cell (length and diameter were
75~mm and 22~mm, respectively) containing isotopically enriched $^{87}$Rb and
30 Torr Ne buffer gas,
mounted inside three-layer magnetic shielding and maintained at the temperature
of $60.5^\circ$C, which corresponds to an optical depth of $\alpha L = 24$. The
laser beam was circularly polarized and weakly focused to $\approx 5$~mm
diameter inside the cell. We found the typical spin wave decay time to be $1/(2
\gamma_\textrm{s}) \simeq 500~\mu$s, most likely arising from small,
uncompensated remnant magnetic fields. The duration of pulses used in the
experiment was short enough for the spin decoherence to have a negligible
effect during writing and retrieval stages and to cause a modest reduction of
the efficiency $\propto \exp{(-2
\gamma_\textrm{s} \tau)}
= 0.82$ during the storage time $\tau~=~100~\mu$s. For the theoretical calculations,
we model the $^{87}$Rb $D_1$ line as a homogeneously broadened \cite{doppler} $\Lambda$-system with no free
parameters, as in Ref.~\cite{novikovaPRLopt} (see Ref.~\cite{phillipsPRA08} for
details).

\begin{figure*}
\includegraphics[width=1.2\columnwidth]{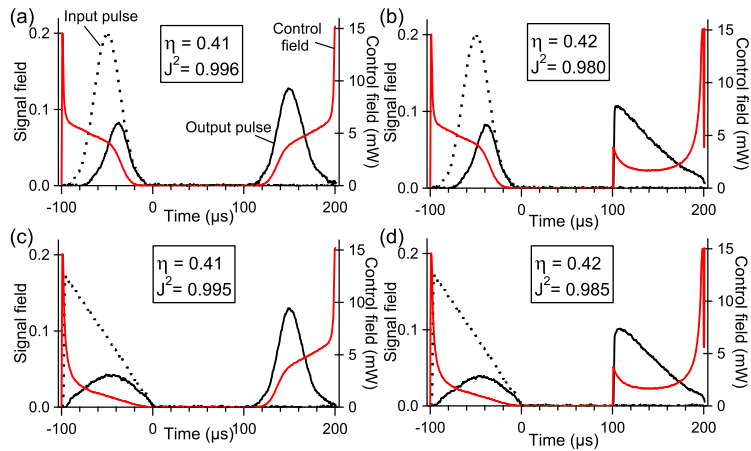}%
\caption {(Color online)  An input Gaussian pulse was optimally stored and retrieved
either into its original pulse shape (a) or into a ramp pulse shape (b).
Similarly, the  incoming ramp pulse was optimally stored and retrieved into a Gaussian (c)
or into an identical ramp (d). Input and output signal pulses are shown as
dotted and solid black lines, respectively, while the optimal control fields
are shown in solid red lines. \label{plug_n_play.fig}}
\end{figure*}
%

An example of optimized 
light storage with controlled retrieval 
 is shown
in Fig.~\ref{basicstorage.fig}(b,c). In this measurement, we chose 
the input pulse $\e_\textrm{in}(t)$~\cite{pulsenote} to be a ``positive
ramp''. According to 
theory~\cite{gorshkovPRL, gorshkovPRA2}, the 
maximum memory efficiency 
is achieved only if the input pulse is mapped onto a particular optimal spin
wave $S(z)$, unique for each $\alpha L$. The calculated optimal spin wave for
$\alpha L=24$ is shown in the inset in Fig.~\ref{basicstorage.fig}(b). Then, we
used the 
method described in Ref.~\cite{gorshkovPRA2} to calculate
the \emph{writing} control field $\Omega(t)$ ($-\mathrm{T}<t<0$) that maps the
incoming pulse onto the optimal spin wave $S(z)$.
%
To calculate the \emph{retrieval} control field $\Omega(t)$
($\tau<t<\tau+\mathrm{T}$) that maps $S(z)$ onto the 
target output pulse $\e_\textrm{tgt}(t)$, we employ the same writing control 
calculation
together with the following time-reversal symmetry of the optimized light
storage \cite{novikovaPRLopt, gorshkovPRL, gorshkovPRA2}. A given input pulse,
stored using its optimal writing control field, is retrieved in the
time-reversed and attenuated copy of itself [$\e_\textrm{out}(t) \propto
\e_\textrm{in}(\tau-t)$] when the time-reversed control is used for retrieval
[$\Omega(t)=\Omega(\tau-t)$]. 
Thus the control field
that retrieves the optimal spin wave
$S(z)$ into $\e_\textrm{tgt}(t)$ 
is the time-reversed copy of the control that stores 
$\e_\textrm{tgt}(\tau - t)$ into $S(z)$.
As shown in Fig.~\ref{basicstorage.fig}(b,c), we used this 
method to achieve 
pulse-shape-preserving storage and retrieval, 
\emph{i.e.},\ the
target output pulse was identical to the input pulse (``positive
ramp''). 
The measured output pulse [solid black line in Fig.~\ref{basicstorage.fig}(c)]
matches very well the target shape [dashed blue line in the same figure]. This
qualitatively demonstrates the effectiveness of the proposed control method.

To describe the memory quantitatively, we define memory efficiency $\eta$ as the probability of retrieving  an incoming 
photon after
some storage time, or, equivalently, as the energy ratio between retrieved and initial signal pulses:
\begin{equation} \label{eta_defin}
\eta=\frac{\int_{\tau}^{\tau+\mathrm{T}}|\e_\textrm{out}(t)|^2dt}{\int_{-\mathrm{T}}^0|\e_\textrm{in}(t)|^2dt}.
\end{equation}
To characterize the quality of pulse shape generation, 
we define an overlap integral $J^2$ as \cite{loudon_book}
\begin{equation} \label{J2_defin}
J^2=\frac{|\int_{\tau}^{\tau+T}\e_\textrm{out}(t)\e_\textrm{tgt}(t)dt|^2}
{\int_{\tau}^{\tau+\mathrm{T}}|\e_\textrm{out}(t)|^2dt
\int_{\tau}^{\tau+T}|\e_\textrm{tgt}(t)|^2dt}.
\end{equation}
The measured memory efficiency for the experiment 
in Fig.~\ref{basicstorage.fig} is $0.42 \pm 0.02$. This value closely approaches
the 
predicted highest achievable 
efficiency $0.45$ for $\alpha L =
24$~\cite{gorshkovPRL, gorshkovPRA2}, corrected to take into account
the spin
wave decay during the storage time. The measured value of the overlap integral
between the output 
and the target 
is $J^2=0.987$, which
indicates little distortion in the retrieved pulse shape.
%

The definitions of efficiency $\eta$ and overlap integral $J^2$ are motivated
by quantum information applications. Storage and retrieval of a single photon
in a non-ideal passive quantum memory produces a mixed state that is
described by a density matrix $\rho = (1-\eta)|0\rangle
\langle0| + \eta
|\phi\rangle \langle \phi|$~\cite{gorshkovPRA1}, where $|\phi\rangle$ is a single photon state with
envelope $\e_\textrm{out}(t)$, and $|0\rangle$ is the vacuum state. Then the fidelity
between the target single-photon state $|\psi\rangle$ with envelope
$\e_\textrm{tgt}(t)$ and the 
single-photon state $|\phi\rangle$ is 
given by the overlap integral $J^2$ [Eq.~(\ref{J2_defin})], while $F = \langle
\psi|\rho|\psi\rangle = \eta J^2$ is 
the 
fidelity of the
output state $\rho$ with respect to the target state $|\psi\rangle$.
The overlap integral $J^2$ is also an essential parameter for optical quantum
computation and communication protocols \cite{DLCZ,kok07}, since $(1-J^2)/2$ is
the coincidence probability in the Hong-Ou-Mandel  \cite{hong87} interference
between photons $|\psi\rangle$ and $|\phi\rangle$~\cite{loudon_book}. 
One should 
be cautious in directly using our classical measurements of $\eta$ and
$J^2$ to predict fidelity for single photon states because single photons may
be sensitive to imperfections that do not significantly affect classical
pulses. For example, 
four-wave mixing processes may reduce the fidelity
of single-photon storage, although our experiments~\cite{phillipsPRA08} found
these effects to be relatively small at $\alpha L < 25$.

%
%

Fig.~\ref{plug_n_play.fig} shows more examples of 
optimal light storage with full output-pulse-shape control. 
For this experiment, we stored either of two randomly selected
input signal pulse shapes --- a Gaussian and a ``negative ramp'' --- and
 then retrieved them either into their original waveforms (a,d) or
into each other (b,c). 
Memory efficiency $\eta$ and overlap integral $J^2$ are
shown for each graph. Notice that the efficiencies 
for all four
input-output combinations are 
very similar ($0.42 \pm 0.02$) and
agree well with the 
highest achievable efficiency ($0.45$) for the given optical depth $\alpha L =
24$. The overlap integrals are also very close to $1$, revealing an excellent
match between the target and the retrieved signal pulse shapes. Note that
different input pulses stored using corresponding (different) optimized writing control
fields but retrieved using identical control fields [pairs (a,c) and (b,d)] 
had identical output envelopes, very close to the target one. This observation,
together with the fact that the measured
memory efficiency is close to the fundamental limit, suggests that indeed 
different initial pulses were mapped onto the same optimal spin wave. This indirectly confirms our control not only over the output signal light field but also over the spin wave.

Our full control over the outgoing wavepacket opens up an interesting
possibility to convert a single photon into a so-called ``time-bin'' qubit ---
a single photon excitation delocalized between two time-resolved wavepackets
(bins). The state of the qubit is encoded in  the relative amplitude and phase
between the two time bins~\cite{brendel99}. Such time-bin qubits are
advantageous for
quantum communication
since they are insensitive to polarization fluctuations and depolarization
during propagation through optical fibers \cite{brendel99}.
We propose to efficiently convert a single photon with an arbitrary envelope into a time-bin
qubit by optimally storing the photon in an atomic ensemble, and then
retrieving it into a time-bin output envelope with well-controlled relative amplitude
and phase using a customized retrieval control field.

\begin{figure*}[t]
\includegraphics[width=1.7\columnwidth]{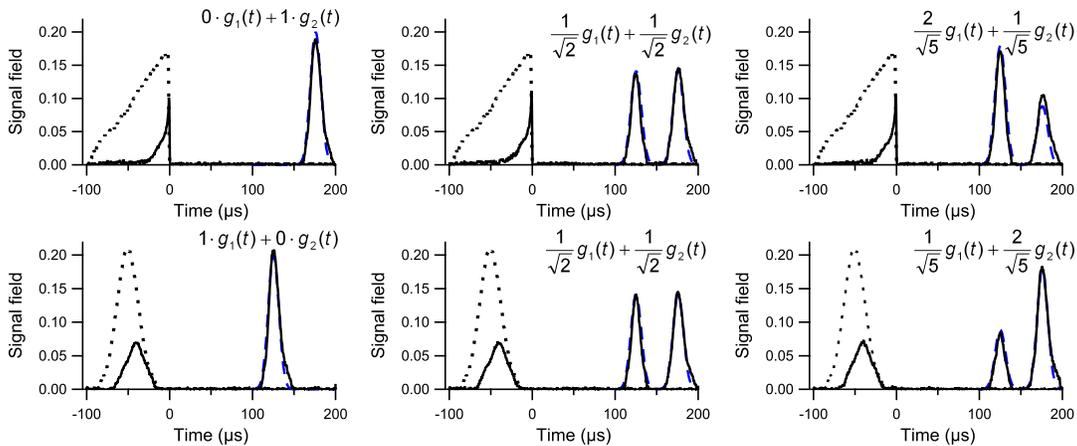}%
\caption {(Color online)  Examples of storage of signal input pulses with
Gaussian and triangular envelopes, followed by retrieval in a 
linear combination of two time-resolved Gaussian pulse shapes $g_1(t)$ and $g_2(t)$.
Input and output signal fields are shown in dotted and solid black lines,
respectively. Dashed blue lines show the target envelopes.
\label{timebin.fig}}
\end{figure*}

To illustrate the proposed wavepacket shaping, in Fig.~\ref{timebin.fig}, we
demonstrate storage of two different input pulses (a Gaussian and a positive
ramp), followed by retrieval into a time-bin-like output pulse, consisting of
two distinct Gaussian wavepackets $g_{1,2}(t)$ with controllable relative
amplitude and delay.
We obtained the target output 
independently of what the input pulse shape was. 
We also attained 
the same memory efficiency as before ($0.41 \pm 0.02$) for all linear
combinations. Also, regardless of the input, the output pulse shapes matched
the target envelopes very well, as characterized by the value of the overlap
integral close to unity $J^2=0.98 \pm 0.01$. We also verified that the
envelopes of the two retrieved components  of the output pulse were nearly
identical by calculating the overlap integral $J^2(g_1,g_2)$ between the retrieved
bins $g_1$ and $g_2$. This parameter is important for applications requiring
interference of the two qubit components~\cite{brendel99}. The average value of
$J^2(g_1,g_2)=0.94 \pm 0.02$ was consistently high across the full range of
target outputs. The relative phase of the two qubit components can be adjusted
by controlling the phase of the control field during retrieval.
The demonstrated control over the amplitude ratio and shape of the two wavepackets is essential
for achieving high-fidelity time-bin qubit generation. Our scheme
is also immediately applicable to high-fidelity partial retrieval of the spin
wave~\cite{hau01}, which forms the basis for a recent promising quantum
communication protocol \cite{sangouard}.


To conclude, we have reported the experimental demonstration of optimal storage
and retrieval of arbitrarily shaped signal pulses in an atomic vapor at an
optical depth $\alpha L=24$ by using customized writing control fields. Our
measured memory efficiency is close to the highest efficiency possible at that
optical depth.  We also demonstrate full precision control over the retrieved
signal pulse shapes, achieved by shaping the retrieval control field.
A high degree of overlap between the retrieved and target pulse shapes was
obtained (overlap integral $J^2=0.98 - 0.99$) for all input and target
pulse shapes tested in the experiments. We also demonstrated the potential application of the presented technique to the creation of optical time-bin qubits and to controlled partial retrieval. Finally, we observed excellent agreement
between our experimental results and theoretical modeling.
The optimal storage and pulse-shape control presented here are
applicable to a wide 
range of experiments, since the underlying theory
applies to other experimentally relevant situations such as ensembles enclosed
in a cavity \cite{simon07, gorshkovPRA1}, the off-resonant regime
\cite{gorshkovPRL, gorshkovPRA1,gorshkovPRA2}, non-adiabatic storage
(\emph{i.e.}, storage of pulses of high bandwidth)~\cite{gorshkovPRA4}, and
ensembles with inhomogeneous broadening \cite{gorshkovPRA3}, including Doppler
broadening \cite{eisaman05} and line broadening in solids \cite{kraus06}. Thus,
we expect this pulse-shape control to be indispensable for applications in both
classical \cite{tucker05} and quantum optical information processing.

The authors would like to thank M. D. Lukin, L. Jiang, A.\ S.\ S{\o}rensen, P.
Walther, and R.\ L.\  Walsworth for useful discussions. This research was
supported by NSF, Jeffress Memorial Trust, and the College of William~\&~Mary.


\begin{thebibliography}{99}
\bibitem{doppler} The homogeneous pressure broadened width
$2 \gamma \approx 2 \pi\times 300$~MHz of the optical transition was comparable to
the Doppler broadening.
\bibitem{pulsenote}
Throughout the paper, all control and signal envelopes are assumed to be real.
Also, in all experimental data plots, all signal pulses are shown in the same
scale, and all input pulses are normalized to have the same unit area
$\int_{-\mathrm{T}}^0 |\e_\textrm{in}(t)|^2dt=1$, where $t$ is time in $\mu$s.
\end{thebibliography}
\end{document}